\documentclass[aip,preprint,amsmath]{revtex4-1}

\draft 

\usepackage{graphicx}
\usepackage{subcaption}
\usepackage{color}
\usepackage{MnSymbol}

\draft 

\definecolor{roig}{rgb}{0,0,0}

\begin{document}


\title{Model selection for radiochromic film dosimetry} 

\author{I. M{\'e}ndez}
\email[]{nmendez@onko-i.si}
\affiliation{Department of Medical Physics, Institute of Oncology Ljubljana, Zalo\v{s}ka cesta 2, Ljubljana 1000, Slovenia}

\begin{abstract}

\textbf{Abstract:}

{\color{roig} The purpose of this study was to find the most accurate model for radiochromic film dosimetry by comparing different channel independent perturbation models. A model selection approach based on (algorithmic) information theory was followed, and the results were validated using gamma-index analysis on a set of benchmark test cases. Several questions were addressed: (a) whether incorporating the information of the non-irradiated film, by scanning prior to irradiation, improves the results; (b) whether lateral corrections are necessary when using multichannel models; (c) whether multichannel dosimetry produces better results than single-channel dosimetry; (d) which multichannel perturbation model provides more accurate film doses. It was found that scanning prior to irradiation and applying lateral corrections improved the accuracy of the results. For some perturbation models, increasing the number of color channels did not result in more accurate film doses. Employing Truncated Normal perturbations was found to provide better results than using Micke-Mayer perturbation models.} Among the models being compared, the triple-channel model with Truncated Normal perturbations, net optical density as the response and subject to the application of lateral corrections was found to be the most accurate model. The scope of this study was circumscribed by the limits under which the models were tested. In this study, the films were irradiated with megavoltage radiotherapy beams, with doses from about 20 cGy to 600 cGy, entire (8 ${\rm inch}$ $\times$ 10 ${\rm inch}$) films were scanned, the functional form of the sensitometric curves was a polynomial and the different lots were calibrated using the plane-based method. 

\end{abstract}

\maketitle 

\section{Introduction}

Near water-equivalence\cite{crijns:2102, aapm:55}, high spatial resolution and weak energy dependence across a broad range of energies\cite{rink:2007, richter:2009, arjomandy:2010, lindsay:2010, massillon:2012, bekerat:2014} make radiochromic film dosimetry with Gafchromic films (Ashland Inc., Wayne, NJ) and flatbed scanners the dosimetry system of choice for many applications in radiation therapy. Radiochromic films darken upon irradiation, which makes it possible to measure the absorbed dose using a scanner. However, when they are digitized with a color scanner, three different dose distributions - one for each color channel - are obtained. To combine the information provided by all three channels into a single and more accurate dose distribution, multichannel dosimetry methods have been proposed\cite{hupe:2006, devic:2009, AMicke:2011, mccaw:2011, mayer:2012, mendez:2013, mendez:2014, perez:2014}.

An emerging field of research in multichannel radiochromic film dosimetry are perturbation models\cite{AMicke:2011, mayer:2012, mendez:2014, perez:2014}. These models consider that, for each channel, the measured dose distribution deviates from the true absorbed dose distribution because of small perturbations in the film-scanner response. To combine all three color channels, certain assumptions on the characteristics of the perturbations are necessary. This approach has shown promising results \cite{AMicke:2011, mayer:2012, vanHoof:2012, mendez:2014, perez:2014}.

The purpose of this study is to select the most accurate model for radiochromic film dosimetry {\color{roig}using Gafchromic EBT2 and EBT3 films and a flatbed scanner}. In order to do so, different channel independent perturbation (CHIP) models are compared. This work examines whether incorporating the information of the non-irradiated film, by scanning prior to irradiation, improves the dosimetry. To what extent perturbation models correct the deviation from the response at the center of the scanner along the axis parallel to the scanner lamp is analyzed. Finally, single-, dual- and triple-channel models are compared to determine if by increasing the number of combined color channels the accuracy of radiochromic film dosimetry is also increased.

The results of this study should be considered applicable within the limits under which the models were tested. Through this work, the films were irradiated with megavoltage radiotherapy beams, in the dose range from about 20 cGy to 600 cGy, and whole (8 ${\rm inch}$ $\times$ 10 ${\rm inch}$) films were scanned. The functional form of the sensitometric curves was a polynomial and the different lots were calibrated using the plan-based method \cite{mendez:2013}.  

\section{Methods and materials}

\subsection{A general perturbation model}

Many disturbances affect the film-scanner response\cite{aapm:55, devic:2006, bouchard:2009, AMicke:2011}: thickness variations in the active layer coated on the film, electronic noise, scanner instability, lateral artifact, local variations produced by systematic problems of the scanner, Newton rings, dust, scratches or other damage, etc. According to their persistence, disturbances can be classified into two groups: systematic local variations and random perturbations. Random perturbations ({\it i.e.}, noise) are not consistent in time: they change or disappear between scan repetitions or between non-irradiated and irradiated film scans. Electronic noise, scanner instability and Newton rings can be included within this category. Systematic variations are consistent between film scans. They include thickness variations in the active layer, lateral artifact, local variations produced by systematic problems of the scanner, etc. Some other disturbances such as dust, scratches or other damage can be random or systematic, but they generally produce large alterations in the response, cannot be treated as perturbations, and should be considered as another source of noise.  
 
The dose absorbed by the film at point $r$, $D(r)$, can be described by:

\begin{equation}
D(r) = D_{k}(z_{k}(r) + \Psi_{k}(r, z_{k})) + \Sigma_{k}(r) 
\end{equation}

where $k$ represents the color channel ({\it i.e.}, red (R), green (G) or blue (B)), $D_{k}$ denotes the sensitometric curve (for channel $k$), $z_{k}(r)$ is the film-scanner response (either pixel value, optical density (OD) or net optical density (NOD)) at point $r$, $\Psi_{k}(r, z_{k})$ corrects the systematic local variations and $\Sigma_{k}(r)$ represents noise disturbances. 

If $\Psi_{k}(r, z_{k})$ is small, we can apply a first-order Taylor expansion of $D(r)$ in terms of $\Psi_{k}(r, z_{k})$. Hence, a general perturbation model for $D(r)$ can be expressed as:

\begin{equation}
\label{general}
D(r) = D_{k}(z_{k}(r)) + \dot{D}_{k}(z_{k}(r))\Psi_{k}(r, z_{k}) + \epsilon_{k}(r) 
\end{equation}

where $\dot{D}_{k}$ is the first derivative of $D_{k}$ with respect to $z_{k}$ and $\epsilon_{k}(r)$ is an error term that accounts for the difference between the true absorbed dose, $D(r)$, and the measured dose after correction of the perturbation. The noise is included in the error term.

\subsection{Channel independent perturbation models in the literature}

\subsubsection{The Micke and Mayer models}

Micke \emph{et al}\cite{AMicke:2011} were the first to suggest a perturbation model. Mayer \emph{et al}\cite{mayer:2012} found a closed-form solution to that model. They proposed a CHIP model:

\begin{equation}
\label{chip}
D(r) = D_{k}(z_{k}(r)) + \dot{D}_{k}(z_{k}(r)) \Delta(r) + \epsilon_{k}(r) 
\end{equation}

which implicitly assumes that the probability density functions (pdf) of all the $\epsilon_{k}(r)$ terms are equal and the pdf of $\Delta$ is a uniform distribution\cite{mendez:2014}.

\subsubsection{The Truncated Normal model}

In an earlier article\cite{mendez:2014}, a generalization of the Micke-Mayer (MM) model was introduced. The pdf of $\Delta$ was considered to be a truncated normal (TN) distribution and the $\epsilon_{k}(r)$ terms could be different. 

In this study, the MM and TN models were compared while different elements of the functional form that translates film-scanner responses into doses were varied. It should be pointed out that the MM model is not the same as the Micke or Mayer multichannel film dosimetry methods. For example, the Mayer method uses pixel value as film-scanner response, by contrast, in this study OD and NOD were used. The Micke method uses a rational form for the sensitometric curves, by contrast, in this study the sensitometric curves followed 

\begin{equation}
\label{sensitometric}
D_{k} = \sum_{n=0}^{4} b_{k_n} z_{k}^n
\end{equation}

where $b_{k}$ are fitting parameters.

However, both the Micke and Mayer methods share the same form of the perturbation, which diverges from the form defined by the TN model.

\subsection{Scanning before and after irradiation}

The response of the dosimetry system to irradiation is expressed in terms of pixel value, OD or NOD. OD is defined as: 

\begin{equation}
z := \log_{10} \frac{v_{\mathrm{max}}}{v_{\mathrm{irr}}}
\end{equation}

where $v_{\rm{irr}}$ denotes the pixel values of the irradiated film and $v_{\rm{max}}$ is the maximum possible pixel value of the scanner. Between pixel value and OD there is only a change of coordinates. For this reason, in this study pixel value was not included in the comparison. 

The information of the non-irradiated film can be incorporated in the response. NOD is defined as:

\begin{equation}
z := \log_{10} \frac{v_{\mathrm{nonirr}}}{v_{\mathrm{irr}}}
\end{equation}

where $v_{\rm{nonirr}}$ denotes the pixel values of the non-irradiated film.

\subsection{The lateral artifact}

The lateral artifact is the deviation from the response at the center of the scanner along the axis parallel to the scanner lamp\cite{fiandra:2006, devic:2006, lynch:2006, Paelinck:2007, Fuss:2007, battum:2008,  Martisikova:2008, menegotti:2008, saur:2008, girard:2012, mendez:2013, poppinga:2014}. It is caused by the interplay between the light scattering from the polymers created in the active layer of the film and the properties of the scanner\cite{Schoenfeld:2014}. It is approximately parabolic in shape, with lower pixel values along the edges than at the center of the scan. It is dependent on the color channel and OD. The lateral correction is modeled empirically. In this paper, lateral corrections are calculated as\cite{saur:2008, mendez:2013}:

\begin{equation}
\label{lateral}
v_{k} = a_{k_1} (x-x_{c}) + a_{k_2} (x-x_{c})^{2} + \hat{v_{k}} 
\end{equation}

where $\hat{v_{k}}$ is the pixel value before correction, $x$ is the coordinate of the pixel in the axis parallel to the CCD array, $x_c$ is the {\it x} coordinate of the center of the scanner, $v_{k}$ represents the corrected pixel value, and $a_{k}$ are fitting parameters.

Multichannel correction methods have been found to substantially mitigate the lateral artifact \cite{AMicke:2011, lewis:2012}. This raises the question of whether additional steps to correct the lateral artifact are necessary. If the lateral correction is considered included in the perturbation term, additional steps are unnecessary. An alternative approach is to explicitly apply the lateral correction following Eq.(\ref{lateral}). In this study, both approaches were compared.

\subsection{Single-channel dosimetry vs. multichannel dosimetry}

Multichannel dosimetry combines the doses measured with different color channels ($D_{k}(r)$) into a single dose value ($d(r)$). Considering $D_{k}(r)$ as different measurements of the dose, if they were uncorrelated the uncertainty of the dose $d(r)$ could not be higher than the uncertainties of each of the channels separately. However, they are correlated \cite{mendez:2014}, which implies that the uncertainty of $d(r)$ can be higher than the uncertainty of $D_{k}(r)$. Thus, multichannel dosimetry is not necessarily better than single-channel dosimetry. It depends, first, on the functional form of the particular multichannel model. According to the functional form, it can be derived that the uncertainty of the CHIP models compared in this work cannot worsen when the number of combined channels increases. Second, this is only valid if the functional form correctly describes the physics of radiochromic film dosimetry.

In this study, single-, dual- and triple-color-channel dosimetry models were compared.

\subsection{Model selection}

The most accurate model for radiochromic film dosimetry will be the one with the highest degree of coincidence between the film dose ($d(r)$) and the true dose ($D(r)$). The degree of coincidence is a qualitative magnitude, in this work it was quantified as the probability of both being equal. 

According to Bayesian probability theory, given the measured data, the probability that a model is the one where the data come from is proportional to the probability of obtaining these data from the model times the prior probability of the model: $ P(M|D) \propto P(D|M) \: P(M)$. The ideal choice for the prior is the universal weight based on the Kolmogorov complexity of the model\cite{hutter:2005, li:2009}. Unfortunately, the Kolmogorov complexity is not finitely computable. A practical complexity-based approach consists of selecting the model using the Akaike information criterion (AIC)\cite{burnham:2002}. Thus, the most probable model is the one with the lowest AIC value: $ AIC = 2\: c_{M} - 2\: \mathrm{ln}(P(D|M))$, where $c_{M}$ is the number of parameters of the model. 

Usually, when selecting a model for film dosimetry, the functional form is predefined and the only parameters to select fit the lateral corrections and sensitometric curves. This is done for each lot of films and is referred to as `calibration'. In this study, the functional form (MM or TN, OD or NOD, etc.) had to be selected as well. Each lot of films was calibrated with each functional form. For each calibration, to calculate $P(D|M)$, it was considered that the differences ($\epsilon_{d}$) between the reference doses ($D(r)$) and measured doses ($d(r)$) were normally distributed. Thus, and considering that the number of parameters was negligible compared to the size of the data sample, the model with the lowest root mean square error (RMSE) comparing the reference and measured doses was regarded as the most likely model. Therefore, given a functional form, the parameters of the model were selected minimizing the RMSE. In addition, for each lot, the most probable functional form was given by the model with the lowest RMSE. 

Henceforth, a given use of the term `model' will refer to all those models with the same functional form. A calibration will select the parameters that maximize the likelihood of that model.

To exclude the possible dependence on a particular lot, a sample with four different lots was employed. Calibrations of different lots had different dose ranges and quantities of data. To balance their weights, relative rather than absolute differences were compared. For each calibration, it was assumed that the relative differences between reference doses ($D(r)$) and measured doses ($d(r)$) were also normally distributed:

\begin{equation}
D(r) = d(r) + \epsilon_{d}(r) 
\end{equation}
 
with:  

\begin{equation}
\label{relative_unc}
\frac{\epsilon_{d}(r)}{d(r)} \sim \mathcal{N}(0,\sigma^{2})  
\end{equation}

where $\sigma$ is referred to as the relative `uncertainty of the calibration'. To estimate the uncertainty of the measured dose, $\sigma$ should be combined with the uncertainty of $D(r)$ . 

In information theory, entropy is defined as the expected negative log-likelihood of a random variable. Hence, maximizing the likelihood of a model is equivalent to minimizing the information entropy of the errors. The (differential) entropy of a model following Eq.(\ref{relative_unc}) can be calculated as:

\begin{equation}
h(M) = \frac{1}{2} \mathrm{ln} (2 \pi \mathrm{e} \sigma^{2})
\end{equation}

Assigning the same weight to each lot, the entropy of a model in a sample of lots is the mean entropy of the model in the sample. As a result, it was considered that the most probable model for radiochromic film dosimetry was the one with the lowest geometric mean of the calibration uncertainty for all lots under study.

\subsection{Scanning protocol}

Gafchromic EBT2 and EBT3 films with dimensions 8 ${\rm inch}$ $\times$ 10 ${\rm inch}$ were used. They were handled in conformity to the recommendations of the AAPM TG-55 report \cite{aapm:55}. The films were scanned with an Epson Expression 10000XL flatbed scanner (Seiko Epson Corporation, Nagano, Japan) prior to irradiation and within the time-window following irradiation. The scanner was warmed up for at least 30 min before use. Before acquisitions, and after long pauses, five empty scans were taken to stabilize the scanner lamp. The films were centered on the scanner with a black opaque cardboard frame. They were scanned in portrait orientation ({\it i.e.}, the short side of the film parallel to the scanner lamp and the long side parallel to the lamp movement axis). Scans were acquired with image-type set to 48-bit RGB (16 bit per channel), a resolution of 72 dpi and image processing tools turned off. They were saved as TIFF files. Five consecutive scans were taken for each film and the first scan was discarded to avoid the warm-up effect of the scanner lamp occurring with multiple scans\cite{Paelinck:2007, Martisikova:2008}. 

\subsection{Calibration}

To calibrate a lot according to the models under comparison, a set of reference doses should be associated with a representative sample of responses, lateral positions and perturbations. 

To obtain a representative calibration sample, the plan-based method was chosen \cite{mendez:2013}. Films were placed in a Plastic Water (Computerized Imaging Reference Systems Inc. Norfolk, VA, USA) phantom at source-axis distance (SAD). They were irradiated with a 6 MV photon beam and a $60^{\circ}$ Enhanced Dynamic Wedge field of dimensions 20$\times$20 ${\rm cm^2}$ at SAD. The range of doses delivered to the film encompassed the range of doses of interest. For some lots, the irradiation was repeated with several films and different monitor units (MU). In this way, the dose range was extended and the intralot variability mitigated. The dose distribution in the plane of the film was either calculated with the treatment planning system (TPS)  or, preferably, measured simultaneously with an IBA MatriXX Evolution ionization chamber array (IBA Dosimetry GmbH, Germany). Since the reference dose plane can be either planned or measured, from now on this method will be referred to as the `plane-based' method.

The plane-based method is an alternative to the well-established calibration method with fragments\cite{mendez:2013}. A disadvantage of the plane-based method is that the reference doses have higher uncertainty than with the calibration with fragments. However, better precision in the reference dose is useless if the calibration sample is biased. The plane-based method provides a more representative calibration sample, making it more robust against perturbations. Additionally, another advantage of the plane-based method is the efficiency, since the time required for calibration is considerably reduced.

Four lots were used, two of them with EBT2 films: lot A04141003BB (Lot A) and lot A03171101A (Lot B); and the other two with EBT3: lot A05151201 (Lot C) and lot A03181301 (Lot D). They were calibrated following slightly different variations of the protocol:

The Lot A films were irradiated at a depth of 6 cm in a 12$\times$30$\times$30 ${\rm cm^3}$ phantom in slab form with a Novalis Tx accelerator (Varian, Palo Alto, CA, USA). One film was irradiated with a wedge field with 660 MU (the reference doses extended from 100 cGy to 600 cGy, approximately). The reference dose plane was calculated with Eclipse v.10.0 (Varian Medical Systems) using the Anisotropic Analytical Algorithm (AAA). The films were scanned in reflection mode and the waiting time-window was 24 $\pm$ 1 h.

The Lot B and Lot C films were irradiated atop the IBA MatriXX detector inside the IBA MatriXX Evolution MULTICube with a Novalis Tx accelerator. Three films were used for the calibration, two of them were irradiated with a wedge field with 535 MU (dose range: 75-400 cGy, approximately) and the other one with a wedge with 401 MU (dose range: 50-300 cGy, approximately). The reference doses were measured simultaneously. The films were scanned in reflection and transmission mode. The time-window was 20 $\pm$ 1 h.

The Lot D films were also irradiated and measured with IBA MatriXX Evolution. They were irradiated with a Varian 2100 CD accelerator. Three films were used for the calibration using wedge fields with 130, 350 and 550 MU (approximate dose ranges: 20-100, 50-270 and 80-420 cGy, respectively). The films were scanned in transmission mode and the time-window was 24 $\pm$ 1 h.

Whenever the reference doses were measured with the detector array, the dose values were scaled with a factor of 1.015 in order to correct for the distance between the plane of the film and the plane of measurement of the detector. When the reference doses were calculated with the TPS, they were corrected with the daily output of the Linac.

The reference dose planes were exported (with a resolution of 0.49 mm/px for the planned doses and 7.62 mm/px for doses measured with MatriXX). They were uploaded together with the film scans to Radiochromic.com (http://radiochromic.com). All the dosimetry models analyzed in this study were incorporated in a research version based on Radiochromic.com v1.6.

A total of six calibration samples ({\it i.e.}, one for each lot except for Lots B and C, which had two samples: one scanned in reflection and another in transmission mode) were uploaded. 

Each was calibrated using MM and TN perturbations, using OD and NOD as film-scanner response, applying and not applying the lateral correction defined in Eq.(\ref{lateral}) and employing each of the seven possible combinations of color channels ({\it i.e.}, R, G, B, RG, RB, GB, RGB). Taking into account that the MM and TN models only apply to multichannel combinations, 44 different models for radiochromic film dosimetry were compared for each of the calibration samples. Since Radiochromic.com optimizes the calibrations using an evolutionary algorithm, each calibration was repeated three times and the one with the lowest uncertainty was selected.

\subsection{Validation}

Given the calibration sample, the model with the lowest calibration uncertainty was considered the most accurate one. This model is expected to correctly describe the relationship between the film-scanner response and the absolute dose absorbed by the film. However, this expectation can be disproved when studying new data, especially if the calibration sample is not representative. To validate the model selection, film dose distributions were compared with planned dose distributions by means of global gamma-index analyses \cite{low:1998}. Tolerances were set at 3\%, 3 mm excluding points with less than 30\% of the maximum planned dose. This threshold was chosen to prevent extrapolating the sensitometric curves of the calibrations for lower doses when using OD. To avoid noise artifacts\cite{Clasie:2012} ({\it i.e.}, noise in the evaluation distribution which spuriously improves the gamma-index), the film and planned distributions were, respectively, the reference and evaluation distributions.

Benchmark test cases described in an earlier paper \cite{mendez:2014} were employed. To allow other researchers to reproduce and verify the results, or to improve the present models under analysis, these benchmark tests are publicly available at Radiochromic.com. They were considered a representative sample of the dose distributions, both for the EBT2 and EBT3 films. The EBT2 films belonged to Lot B and the EBT3 films to Lot C. The films were exposed with a Novalis Tx accelerator and scanned, in reflection and transmission modes, following the same protocol as for the calibrations. Three phantoms were used: IBA MatriXX Evolution MULTICube, CIRS Pelvic Phantom (Model 002PRA, Computerized Imaging Reference Systems Inc. Norfolk, VA, USA) and CIRS Thorax Phantom (Model 002LFC). They were set up at SAD. The films were placed atop the MatriXX Evolution detector (coronal plane) in the MatriXX phantom and with an offset from the beam axis in the CIRS phantoms (transversal plane), of 1.5 cm in the thorax phantom and 1.3 cm in the pelvic phantom, to diminish beam attenuation by the film\cite{Kunzler:2009, Arjomandy:2012}. Calibration and verification tests which were simultaneously measured with MatriXX were compared with the TPS, obtaining a mean $\gamma_{<1}$ (3\% 3mm) of 98.9\%.

The film scans (42 tests) were converted to dose distributions applying each of the 44 dosimetry models of the calibration. Radiochromic.com was employed for the calculations. 

\section{Results and discussion}

\begin{figure}
\includegraphics[width=\linewidth]{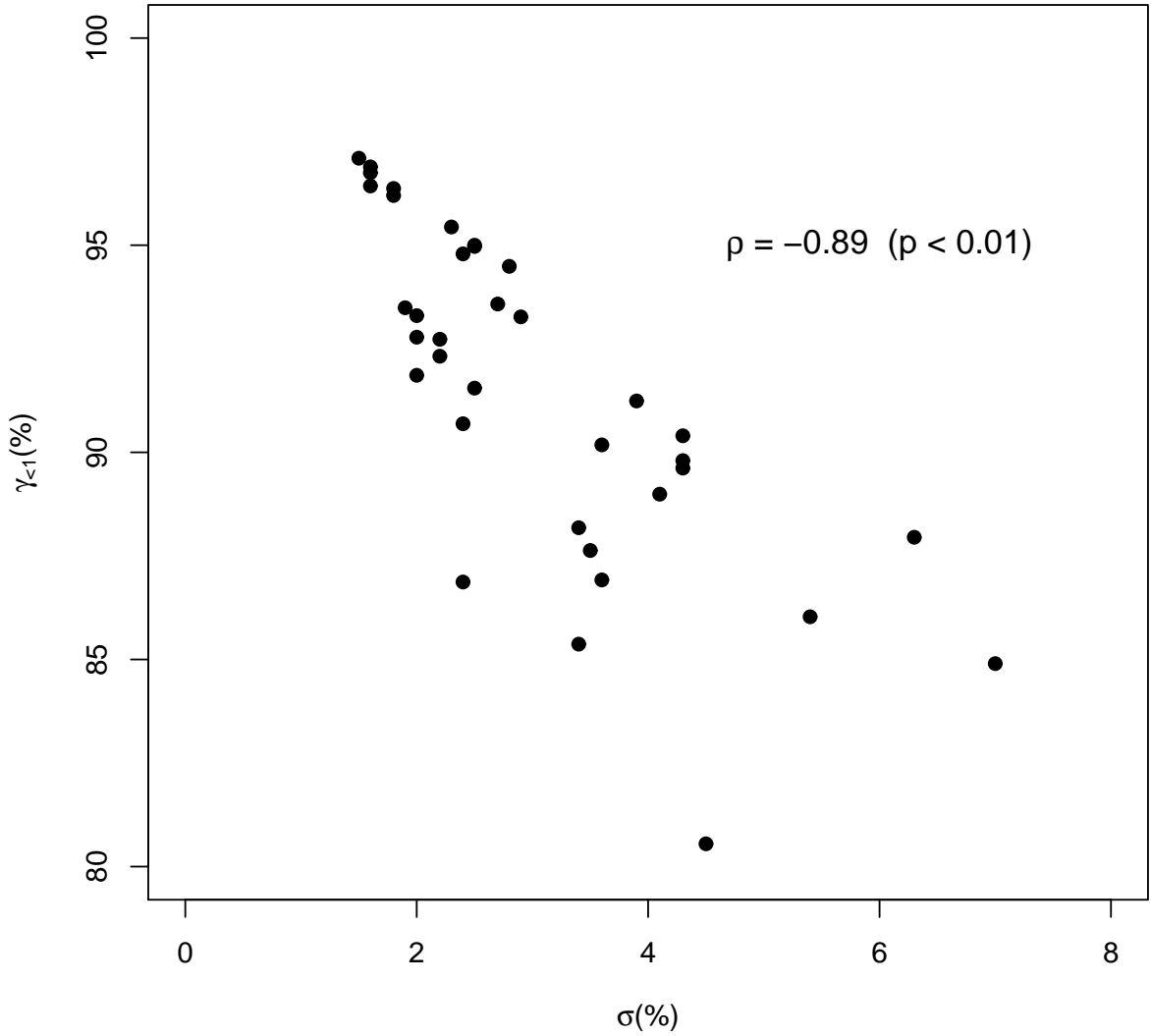}
\caption{\label{fig:correlation} Correlation between the uncertainty of the calibration and the mean, aggregating all 42 test cases, $\gamma_{<1}$ (3\% 3mm). The Spearman's rank correlation coefficient and $p$-value are included}
\end{figure}

If the model selection based on the calibration is correct, the uncertainty of the calibration ($\sigma$) should be correlated with the percentage of points with $\gamma_{<1}$ in the gamma analysis. In {\color{roig}Figure}~\ref{fig:correlation}, it is found that the correlation between $\sigma$ and the mean, aggregating all 42 test cases, $\gamma_{<1}$ (3\% 3mm) is significant. 

In this section, the 44 models are compared. Each dosimetry model comprises: perturbation form for multichannel models (MM/TN), response (OD/NOD), lateral correction (applied/not applied) and the combination of color channels (R, G, B, RG, RB, GB, RGB). Models that only differ in one of these elements will be contrasted according to the $\sigma$ and the mean $\gamma_{<1}$ (3\% 3mm). For the sake of clarity of the presentation, points with $\sigma$ larger than 10\% or $\gamma_{<1}$ less than 80\% are not displayed in the figures.   

\subsection{MM vs. TN perturbations}

\begin{figure*}
\begin{minipage}[b]{0.47\linewidth}
\centering
\includegraphics[width=\linewidth]{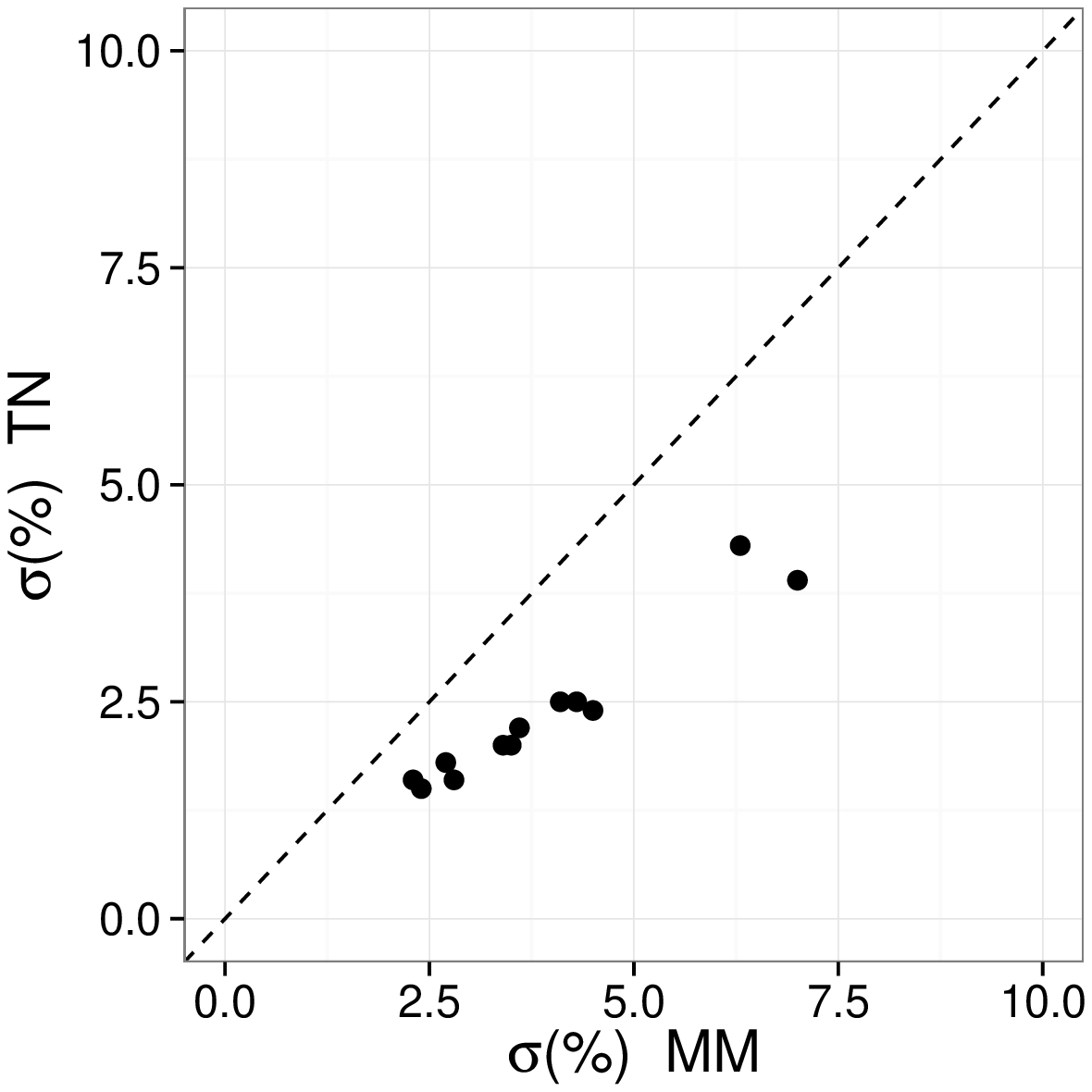}\\
(a)
\end{minipage}
\hfill
\begin{minipage}[b]{0.47\linewidth}
\centering
\includegraphics[width=\linewidth]{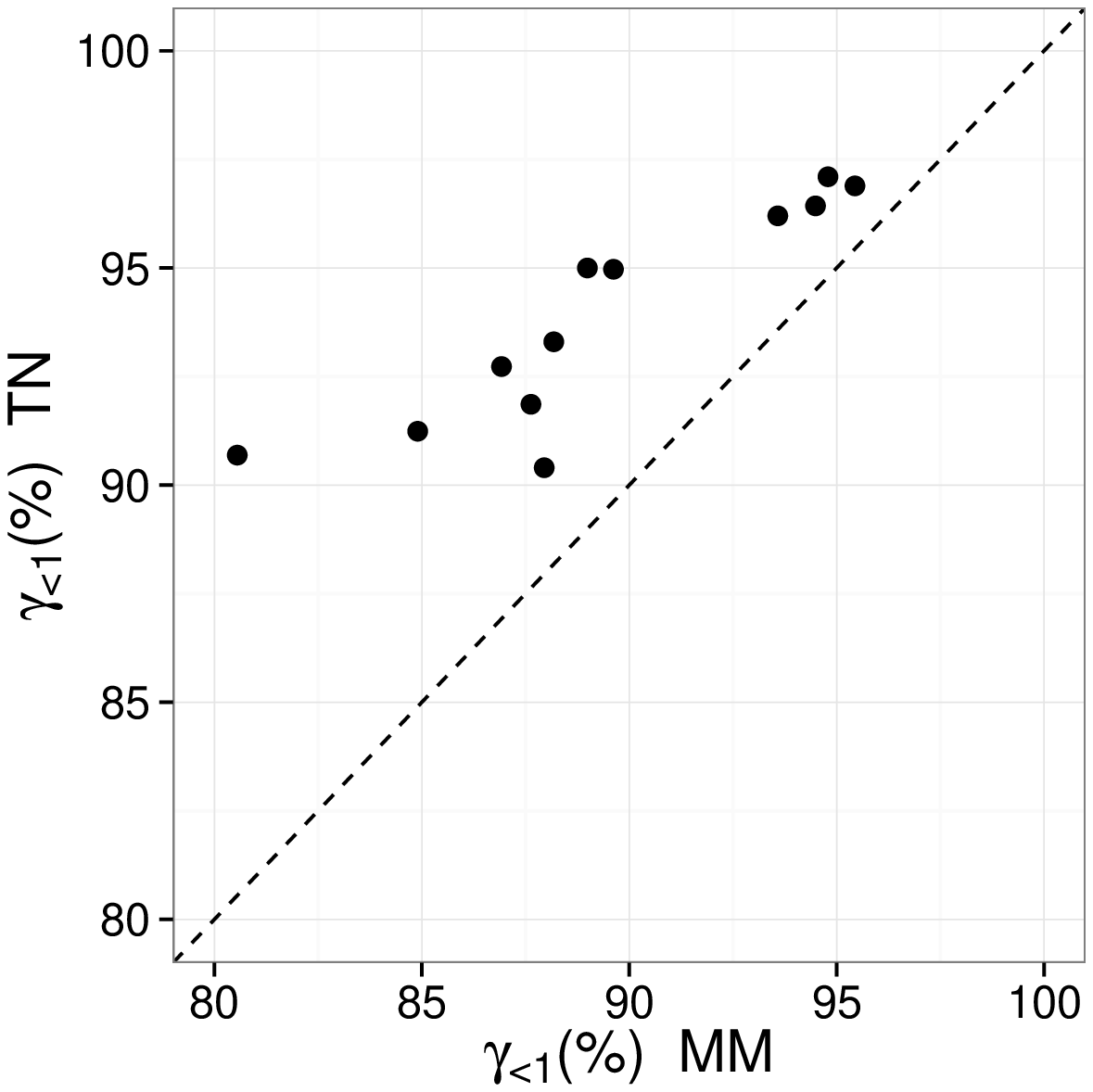}\\
(b)
\end{minipage}
\caption{\label{fig:mm_tn} Contrast of multichannel models that only differ in the form of the perturbation: MM vs. TN perturbations. According to a) the uncertainty of the calibration and b) the mean $\gamma_{<1}$ (3\% 3mm)}
\end{figure*}

{\color{roig}Figure}~\ref{fig:mm_tn} contrasts multichannel models that only differ in the form of the perturbation: they use either MM or TN perturbations. The TN models were found to be better than the MM models in every case. This confirms the recommendation of using TN perturbations as opposed to MM perturbations when using CHIP models. It is due to the fact that the TN perturbation generalizes MM and minimizes the uncertainty in the dose inherent to CHIP models\cite{mendez:2014}.

\subsection{Lateral correction}

\begin{figure*}
\begin{minipage}[b]{0.47\linewidth}
\centering
\includegraphics[width=\linewidth]{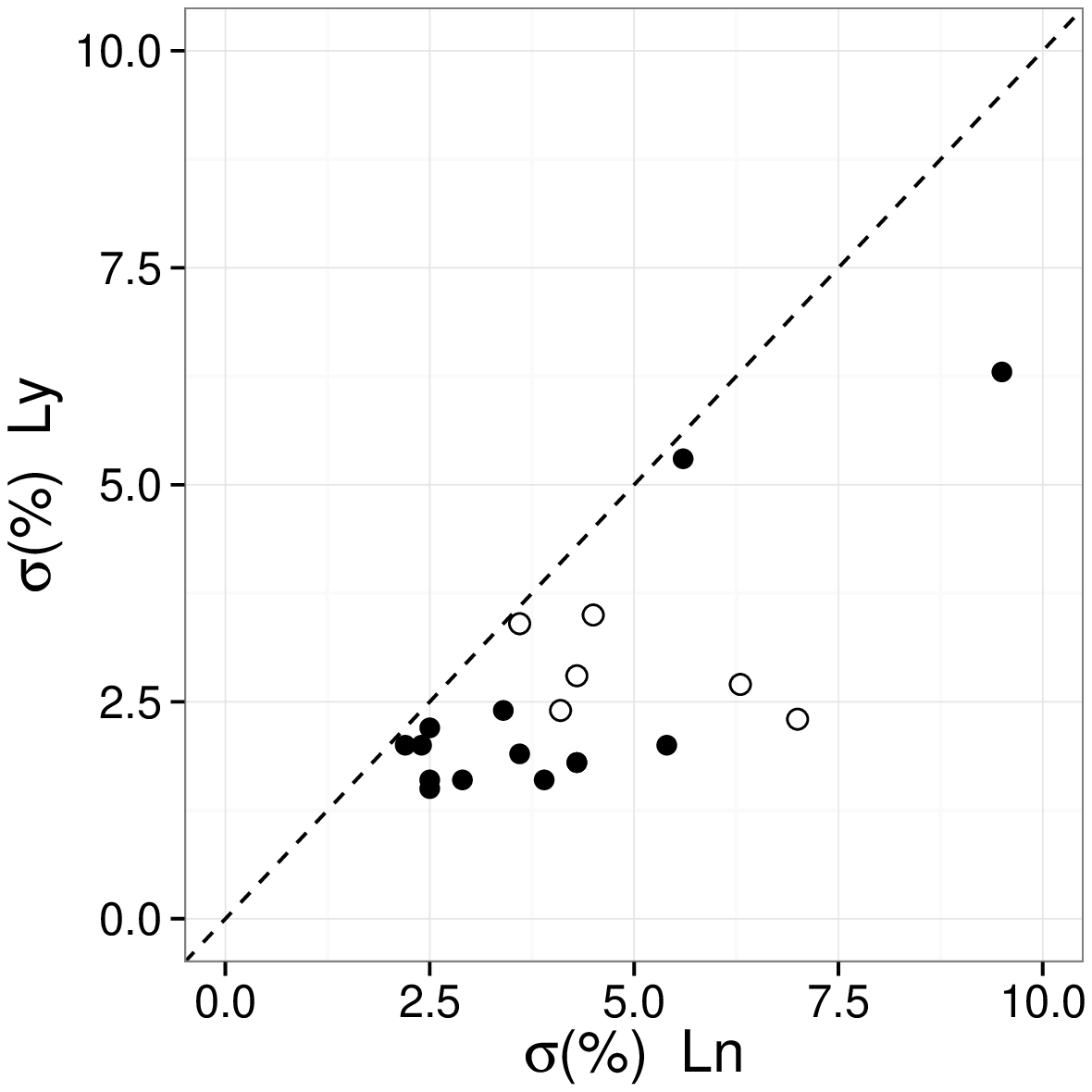}\\
(a)
\end{minipage}
\hfill
\begin{minipage}[b]{0.47\linewidth}
\centering
\includegraphics[width=\linewidth]{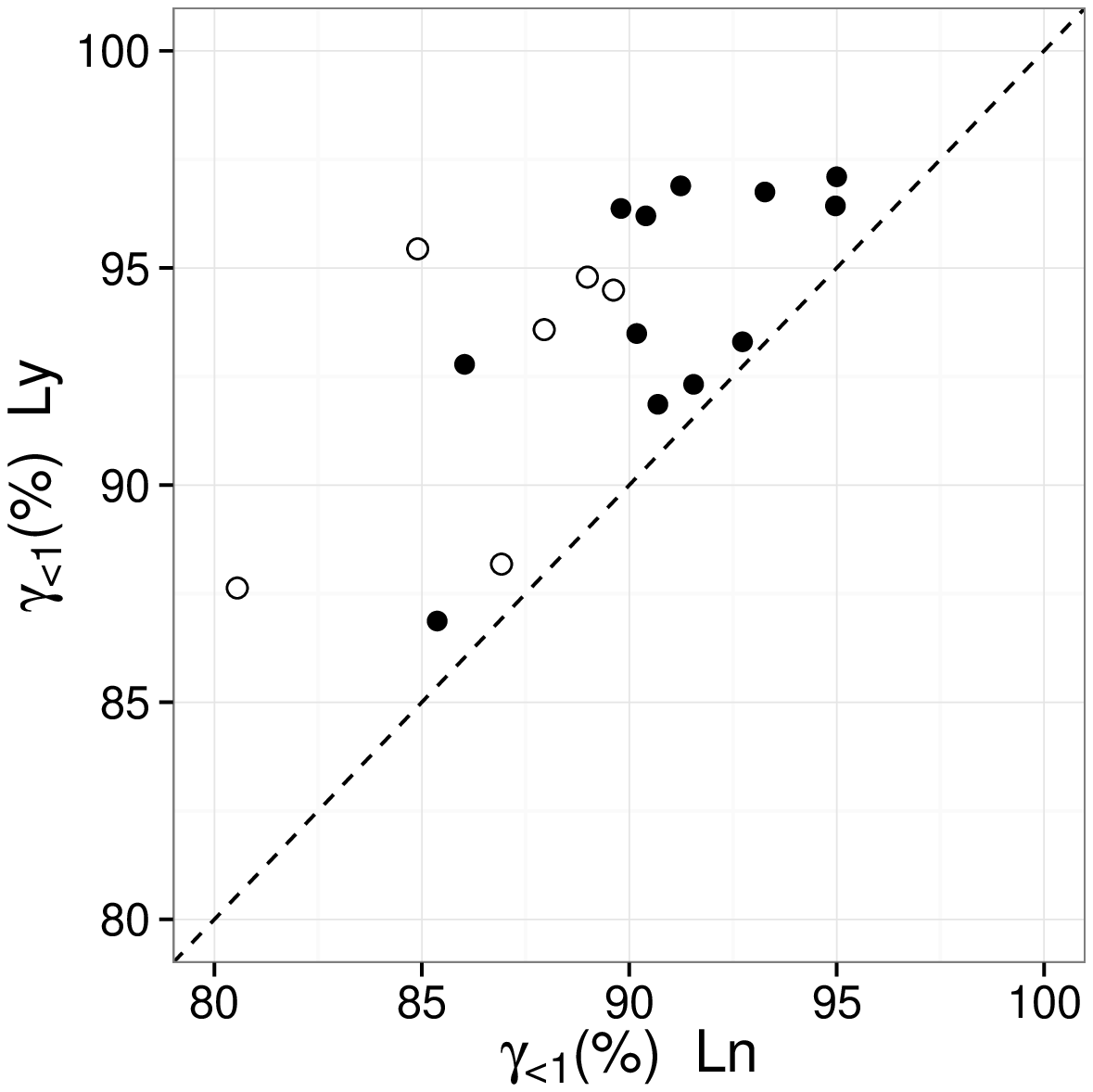}\\
(b)
\end{minipage}
\caption{\label{fig:lateral} Contrast of models that apply (Ly)/do not apply (Ln) lateral corrections. According to a) the uncertainty of the calibration and b) the mean $\gamma_{<1}$ (3\% 3mm). White points are models using MM perturbations.}
\end{figure*}

{\color{roig}Figure}~\ref{fig:lateral} contrasts the models that apply (Ly) with those that do not apply (Ln) lateral corrections. Applying lateral corrections improved the results for all the cases, even for multichannel models.

\begin{figure*}
\begin{minipage}[b]{0.32\linewidth}
\centering
\includegraphics[width=\linewidth]{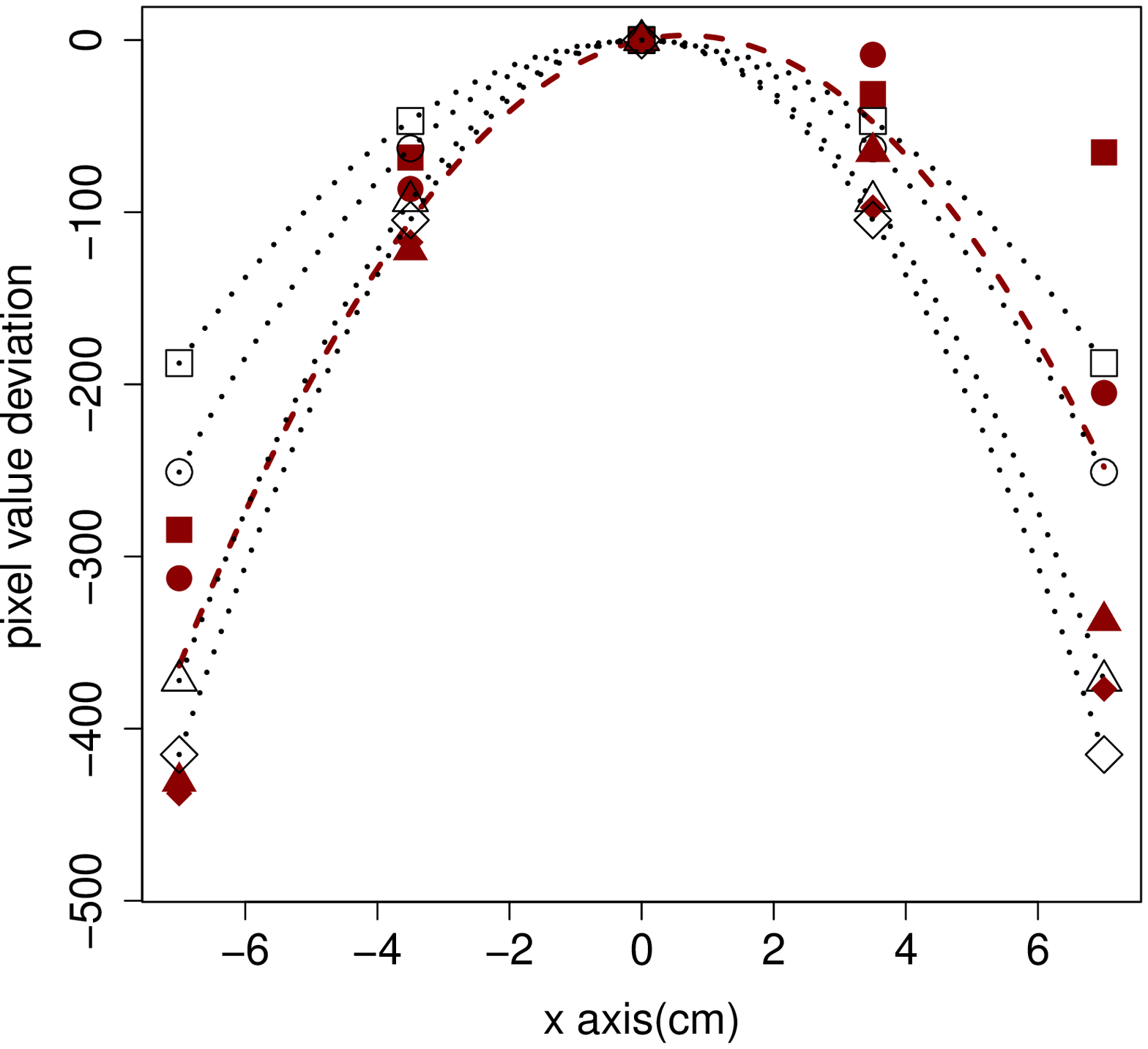}\\
(R)
\end{minipage}
\hfill
\begin{minipage}[b]{0.32\linewidth}
\centering
\includegraphics[width=\linewidth]{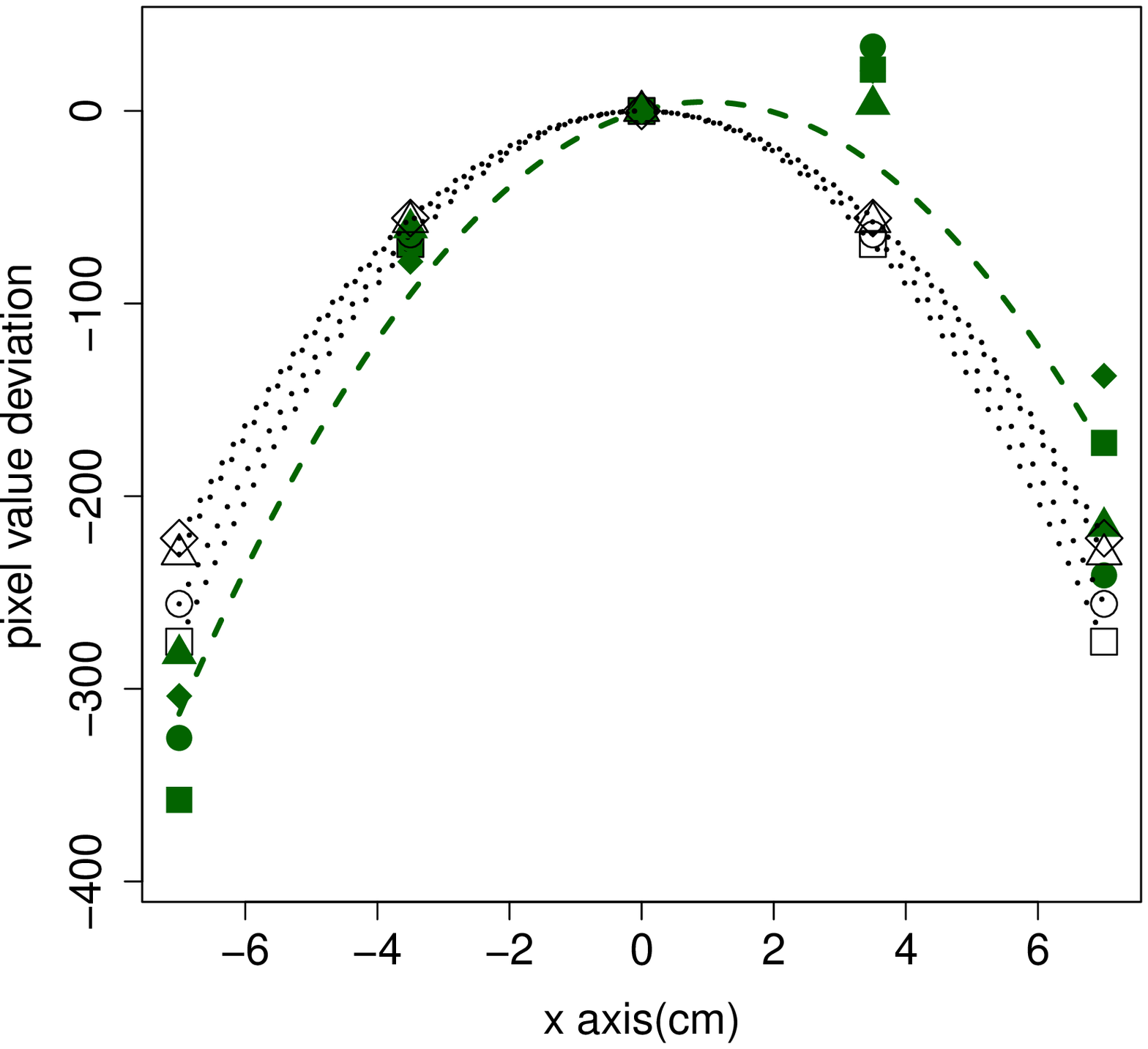}\\
(G)
\end{minipage}
\hfill
\begin{minipage}[b]{0.32\linewidth}
\centering
\includegraphics[width=\linewidth]{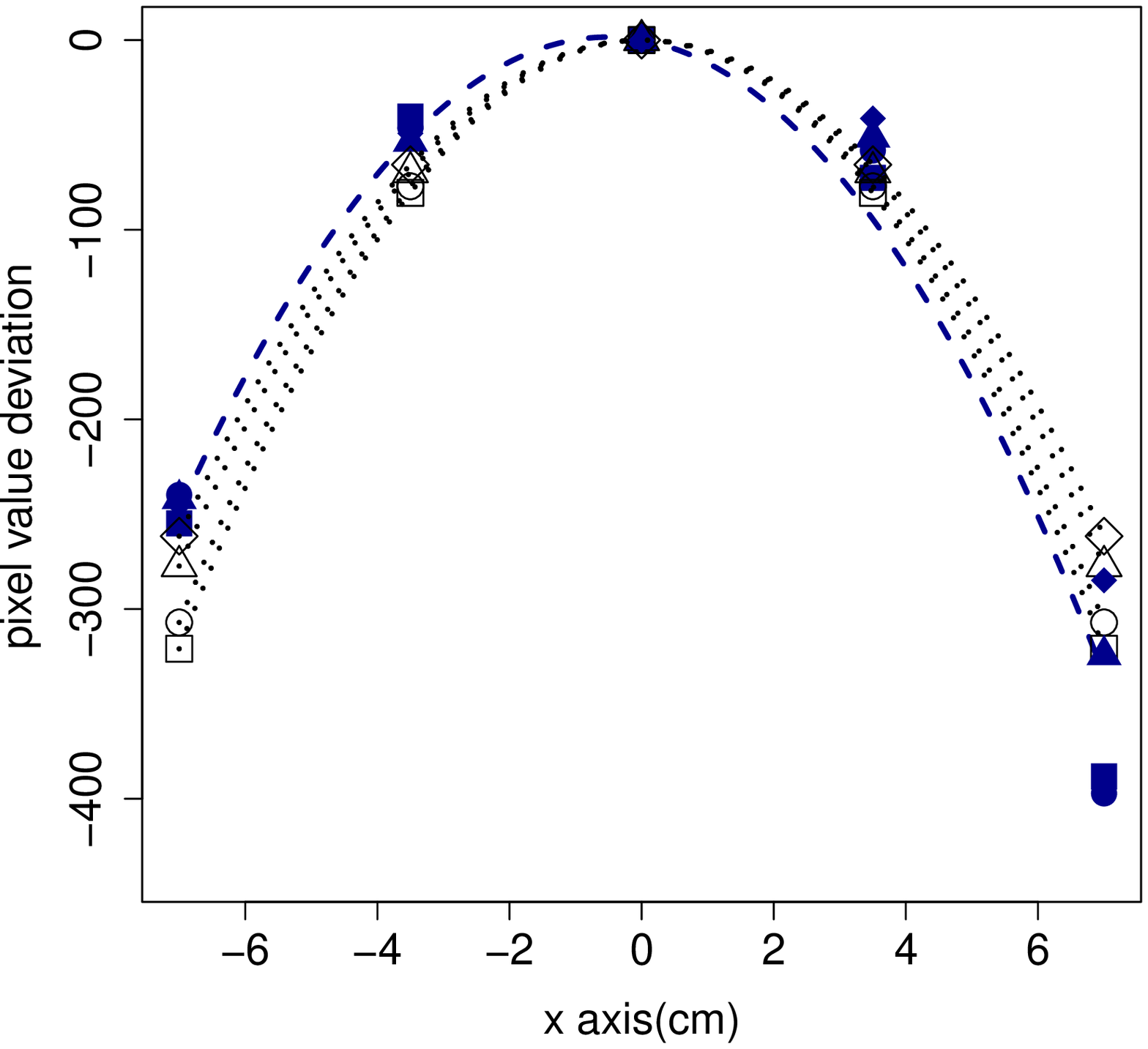}\\
(B)
\end{minipage}\caption{\label{fig:lateral_v}  Film fragments scanned at different positions along the axis parallel to the scanner lamp for the three color channels. Deviations in pixel value from the value measured at the center of the scanner are shown. The colored points represent measurements, the dashed lines Absolute corrections, and the dotted lines and empty points Poppinga corrections. The shape of the point identifies the dose level of the fragment: 0 Gy ($\square$), 1 Gy ($\medcircle$), 4 Gy ($\triangle$), 6 Gy ($\Diamond$).}
\end{figure*}

Recently, a new empirical formula to correct the lateral artifact has been published by Poppinga \emph{et al}\cite{poppinga:2014}. {\color{roig} Figure}~\ref{fig:lateral_v} illustrates how this formula compares with the correction used in this work (Eq.(\ref{lateral})), which will be denoted as Absolute correction.

Four fragments from lot A irradiated with different doses were scanned at five different positions along the axis parallel to the scanner lamp. A 50 $\times$ 50 px ROI with resolution 72dpi was measured at the center of each fragment. The lateral artifacts for all three color channels were fitted following the Absolute correction and the Poppinga correction formulas. {\color{roig}In Figure}~\ref{fig:lateral_v} deviations in pixel value from the value measured at the center of the scanner are plotted. The maximum percentage differences between OD obtained using both correction methods were 0.9\%, 0.3\% and 0.3\% for the R, G and B channels, respectively.

To evaluate both corrections, residuals of the fits were calculated, which are shown in Table~\ref{tab:lateral}. The Poppinga formula was better for the R channel, and the Absolute formula for the G and B channels. It was considered that both lateral corrections were equally valid for empirically modeling the lateral artifact present in radiochromic film dosimetry.

\begin{table}
\caption{\label{tab:lateral} Residuals fitting lateral artifacts with the Absolute and Poppinga formulas for all three color channels.}
\begin{ruledtabular}
\begin{tabular}{lccc}
& \multicolumn{3}{c}{Residuals (pixel value)} \\
\cline{2-4}
Lateral correction & R & G & B\\
\hline
Absolute & 65.3 & 33.7 & 28.4\\
Poppinga & 46.3 & 57.9 & 38.9\\
\end{tabular}
\end{ruledtabular}
\end{table}

\subsection{Scanning before and after irradiation}

\begin{figure*}
\begin{minipage}[b]{0.47\linewidth}
\centering
\includegraphics[width=\linewidth]{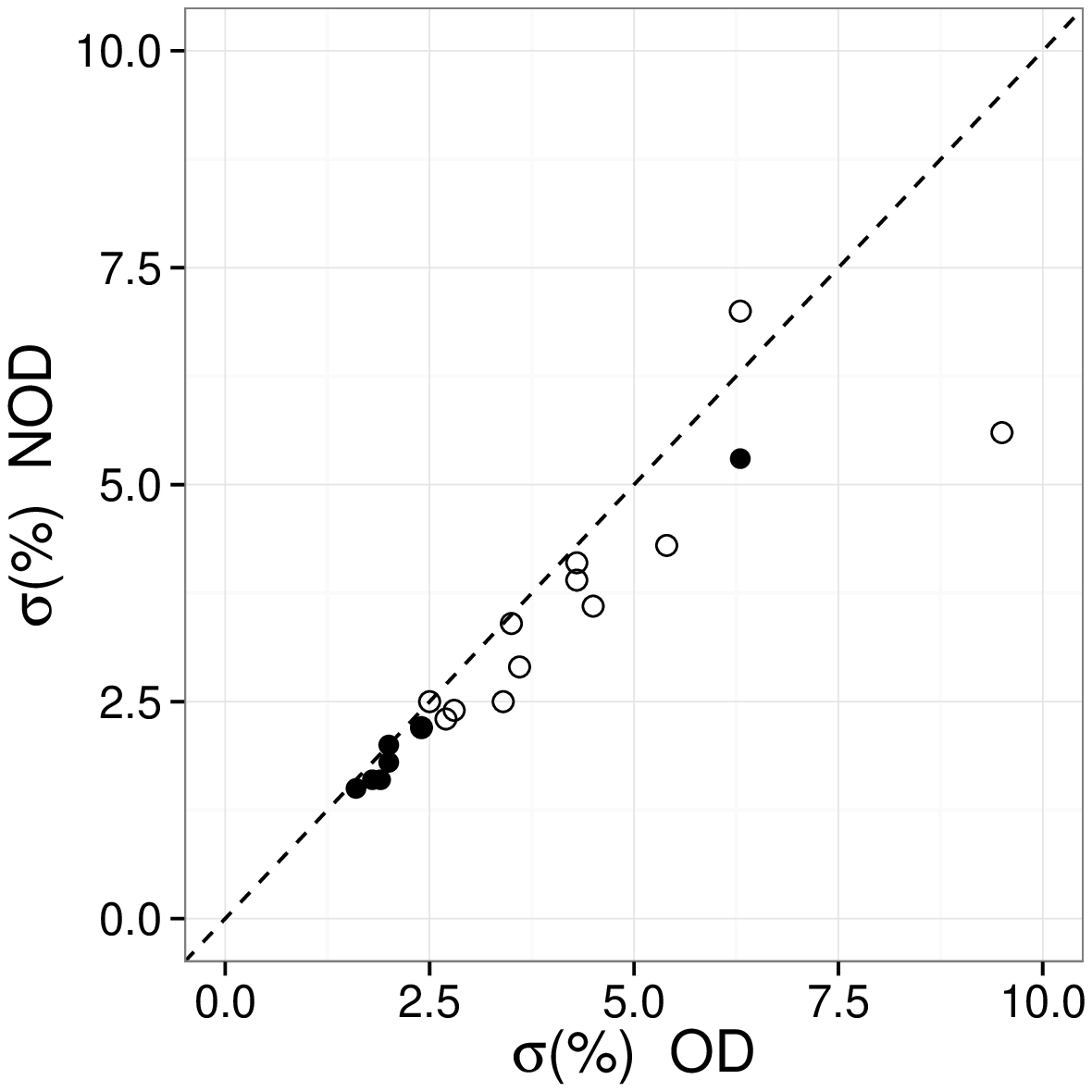}\\
(a)
\end{minipage}
\hfill
\begin{minipage}[b]{0.47\linewidth}
\centering
\includegraphics[width=\linewidth]{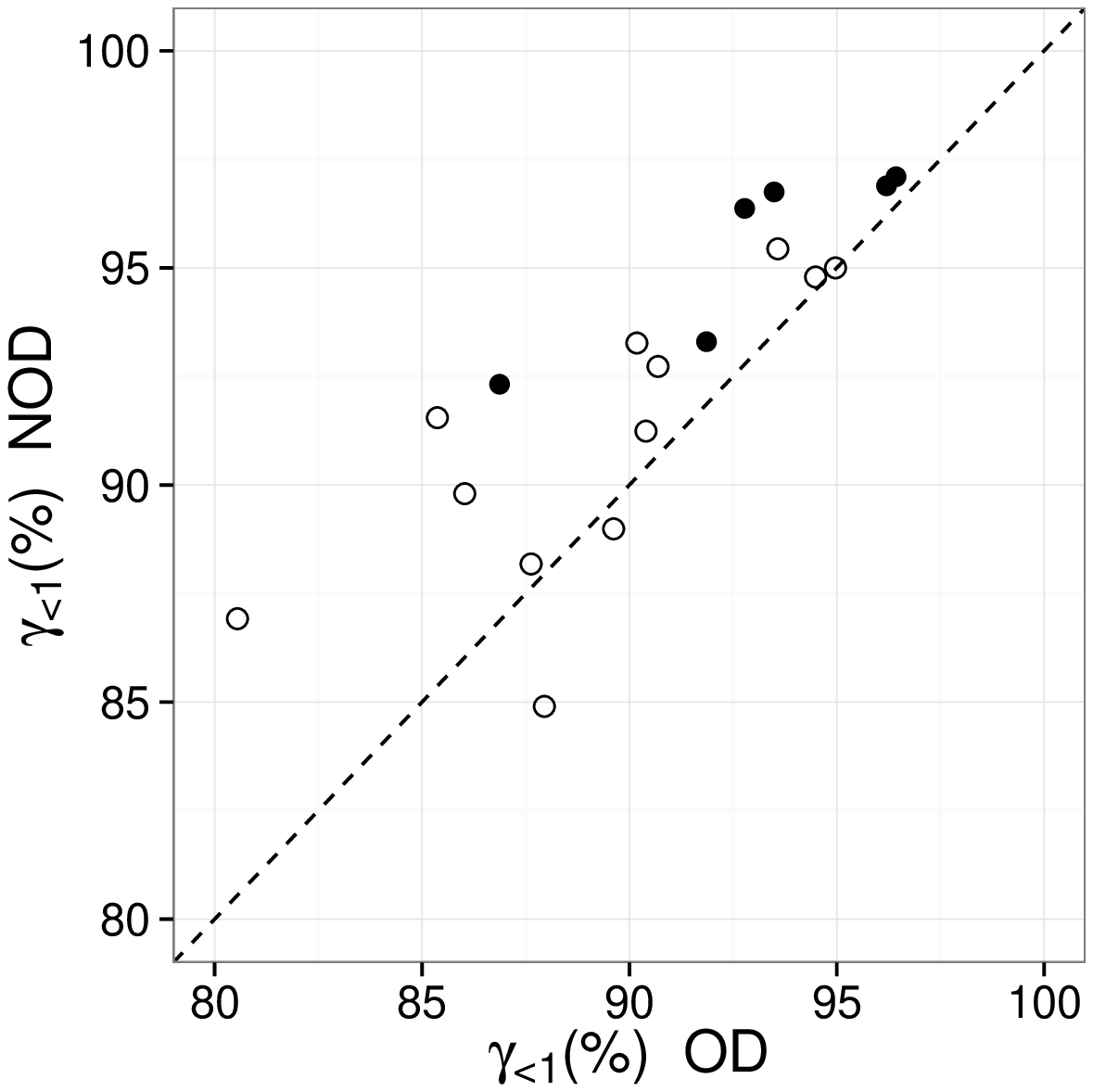}\\
(b)
\end{minipage}

\caption{\label{fig:response} Contrast of models using OD vs. NOD as film-scanner response, according to the uncertainty of the calibration, and according to the mean $\gamma_{<1}$ (3\% 3mm). White points are models that do not apply lateral corrections or use MM perturbations.}
\end{figure*}

{\color{roig}Figure}~\ref{fig:response} contrasts models that only need the information of the irradiated scan ({\it i.e.}, use OD as response) with models that also need the information of the non-irradiated scan ({\it i.e.}, use NOD as response). White points are models that do not apply lateral corrections or use MM perturbations.

Incorporating the information of the non-irradiated scan was found to be correlated with lower $\sigma$ and higher $\gamma_{<1}$ (3\% 3mm). 

When using NOD, dose-independent perturbations in the film-scanner response (e.g., thickness variations of non-active layers) are canceled out, dose-dependent perturbations, like the lateral artifact or film inhomogeneities, are reduced. However, this is valid if the non-irradiated and the irradiated film images are perfectly registered. Registration errors introduce another source of noise/uncertainty that will increase in importance the steeper the gradients of the perturbations are. Registration errors and pixel value noise could increase the noise in the film dose distributions. Even though higher noise in the reference dose distribution is not expected to improve the results of the gamma analysis, in contrast to higher noise in the evaluation distribution, the effect of increasing the noise in the film dose distribution was studied. The dose distribution of a RapidArc prostate plan (case J from the benchmark tests)\cite{mendez:2014} was calculated using the TN-NOD-Ly-RGB model (comprising TN perturbation, with NOD as response, applying lateral correction, and employing all three color channels). The film belonged to Lot C scanned in reflection mode. The noise in the film dose map was either smoothed using median filters or increased with Gaussian noise. The percentage of points with $\gamma_{<1}$ was calculated using tolerances of 3\% 3mm and 2\% 2mm. Data are shown in Table~\ref{tab:noise}. It was found that higher noise in the dose distribution did not improve the gamma results.

\begin{table}
\caption{\label{tab:noise} Percentage of points with $\gamma_{<1}$ (2\% 2mm) and $\gamma_{<1}$ (3\% 3mm) for the test J of Lot C scanned in reflection mode after filtering the dose distribution with median filters or Gaussian noise.}
\begin{ruledtabular}
\begin{tabular}{lcc}
Filter & $\gamma_{<1}$(2\% 2mm) & $\gamma_{<1}$(3\% 3mm)\\
\hline
Median filter 7x7 px & 97.2 & 99.8\\
Median filter 5x5 px & 97.1 & 99.8\\
Median filter 3x3 px & 96.9 & 99.8\\
No filter & 96.0 & 99.6\\
1\% standard deviation noise & 95.1 & 99.3\\
2\% standard deviation noise & 92.6 & 98.3\\
3\% standard deviation noise & 88.7 & 96.8\\
\end{tabular}
\end{ruledtabular}
\end{table}
	
\subsection{Single-channel dosimetry vs. multichannel dosimetry}

Table~\ref{tab:cal_multichannel} and Table~\ref{tab:gam_multichannel} include, respectively, the uncertainty of the calibration and the mean percentage of points with $\gamma_{<1}$ (3\% 3mm), taking into account all 42 test cases, for the dosimetry models under comparison. It can be observed that increasing the number of color channels does not necessarily result in more accurate film doses. This is conspicuous for the MM-RG models, which have very high uncertainties and lower mean $\gamma_{<1}$ (3\% 3mm) values than the corresponding R single-channel models. These high uncertainties and lower mean $\gamma_{<1}$ values were caused by the sensitivity of the inherent dose uncertainty in the MM model with the properties of the sensitometric curves\cite{mendez:2014}; in particular, in these cases, poor outcomes occurred because the sensitometric curves of both the R and G channels were very similar. 

From the data presented in Table~\ref{tab:cal_multichannel} and Table~\ref{tab:gam_multichannel}, it is found that applying lateral corrections and scanning prior and following to irradiation had a larger effect on the results than using multichannel models. For instance, the OD-Ln-R model had 5.4\% of calibration uncertainty and the mean percentage of points with $\gamma_{<1}$ 3\% 3mm was 86\%, the NOD-Ly-R model had 1.8\% and 96.4\% of calibration uncertainty and mean $\gamma_{<1}$ 3\% 3mm, respectively, a greater improvement than the one obtained with the MM-OD-Ln-RGB (4.3\%, 89.6\%) or the TN-OD-Ln-RGB model (2.5\%, 95.0\%).

\begin{table}
\caption{\label{tab:cal_multichannel} Uncertainty of the calibration for the dosimetry models under comparison.}
\begin{ruledtabular}
\begin{tabular}{lccccccc}
& \multicolumn{7}{c}{Color channel combination} \\
\cline{2-8}Dosimetry model & R & G & RG & B & RB & GB & RGB\\
\hline
MM-OD-Ln  & 5.4& 3.4& 81.7&	9.5& 6.3& 4.5& 4.3 \\
MM-OD-Ly  & 2.0& 2.4& 20.6&	6.3& 2.7& 3.5& 2.8 \\
MM-NOD-Ln & 4.3& 2.5& 62.8& 5.6& 7.0& 3.6& 4.1 \\
MM-NOD-Ly & 1.8& 2.2& 14.7&	5.3& 2.3& 3.4& 2.4 \\
TN-OD-Ln  & 5.4& 3.4& 3.6&  9.5& 4.3& 2.4& 2.5 \\
TN-OD-Ly  & 2.0& 2.4& 1.9&  6.3& 1.8& 2.0& 1.6 \\
TN-NOD-Ln & 4.3& 2.5& 2.9&  5.6& 3.9& 2.2& 2.5 \\
TN-NOD-Ly & 1.8& 2.2& 1.6&  5.3& 1.6& 2.0& 1.5 \\
\end{tabular}
\end{ruledtabular}
\end{table}

\begin{table}
\caption{\label{tab:gam_multichannel} Mean percentage of points, taking into account all 42 test cases, with $\gamma_{<1}$ (3\% 3mm) for the dosimetry models under comparison.}
\begin{ruledtabular}
\begin{tabular}{lccccccc}
& \multicolumn{7}{c}{Color channel combination} \\
\cline{2-8}Dosimetry model & R & G & RG & B & RB & GB & RGB\\
\hline
MM-OD-Ln  & 86.0& 85.4& 59.5& 59.2& 88.0& 80.6& 89.6 \\
MM-OD-Ly  & 92.8& 86.9& 72.1& 64.6& 93.6& 87.6& 94.5 \\
MM-NOD-Ln & 89.9& 91.6& 59.6& 76.5& 84.9& 86.9& 89.0 \\
MM-NOD-Ly & 96.4& 92.3& 75.3& 76.2& 95.4& 88.2& 94.8 \\
TN-OD-Ln  & 86.0& 85.4& 90.2& 59.2& 90.4& 90.7& 95.0 \\
TN-OD-Ly  & 92.8& 86.9& 93.5& 64.6& 96.2& 91.9& 96.4 \\
TN-NOD-Ln & 89.9& 91.6& 93.3& 76.5& 91.2& 92.7& 95.0 \\
TN-NOD-Ly & 96.4& 92.3& 96.8& 76.2& 96.9& 93.3& 97.1 \\
\end{tabular}
\end{ruledtabular}
\end{table}

\begin{table}
\caption{\label{tab:thebest} The best models according to: calibration uncertainty ($\sigma$), mean percentage of points with $\gamma_{<1}$ 2\% 2mm (mean $\gamma_{22}$) and 3\% 3mm (mean $\gamma_{33}$), both gamma dimensions taking into account the 42 test cases. They are sorted by mean $\gamma_{33}$.}
\begin{ruledtabular}
\begin{tabular}{lccc}
Dosimetry model & $\sigma$ (\%) & mean $\gamma_{22}$ (\%) & mean $\gamma_{33}$ (\%)\\
\hline
TN-NOD-Ly-RGB & 1.5 & 	88.9 &	97.1 \\
TN-NOD-Ly-RB & 	1.6 & 	88.1 &	96.9 \\
TN-NOD-Ly-RG & 	1.6 & 	88.6 &	96.8 \\
TN-OD-Ly-RGB & 	1.6 & 	87.2 &	96.4 \\
NOD-Ly-R & 		1.8 & 	87.6 &	96.4 \\
TN-OD-Ly-RB & 	1.8 & 	86.4 &	96.2 \\
\end{tabular}
\end{ruledtabular}
\end{table}

Both taking into account the calibration uncertainty and the mean percentage of points with $\gamma_{<1}$ (3\% 3mm), the dosimetry model comprising TN perturbation, with NOD as response, applying the lateral correction, and employing all three color channels ({\it i.e.}, the TN-NOD-Ly-RGB model) obtained the best results and was regarded as the most accurate model in the study.

The models which were among the ten best according to each of these dimensions: calibration uncertainty ($\sigma$), mean percentage of points with $\gamma_{<1}$ 2\% 2mm (denoted as mean $\gamma_{22}$) and 3\% 3mm (mean $\gamma_{33}$), both gamma dimensions taking into account the 42 test cases, are shown in Table~\ref{tab:thebest}. According to these criteria, there were six models. All of them use lateral corrections. Five of them are multichannel models, the other one is a single-channel (R) model. All the multichannel models use TN perturbations. Four out of six models incorporate the information of the non-irradiated film scan. All these models include the R color channel, which is the channel with the highest absorption\cite{devic:07}. 

\subsection{Statistical hypothesis testing}

\begin{table}
\caption{\label{tab:inference} Comparison of models using paired differences of $\gamma_{<1}$(3\% 3mm) values. The sample is composed of the 42 test cases. Median $\gamma_{<1}$ values are shown. Positive values indicate that model A obtained better results, and the opposite for negative values. $P$-values using the Wilcoxon signed-rank test are included.}
\begin{ruledtabular}
\begin{tabular}{cccc}
Model A & Model B & $\gamma_{<1}^{A} - \gamma_{<1}^{B}$ & $p$-value \\
\hline
OD-Ln-R & OD-Ly-R & -4.47 & \textless 0.01 \\
OD-Ln-R & NOD-Ln-R & -3.05 & \textless 0.01 \\
OD-Ln-R & NOD-Ly-R & -7.67 & \textless 0.01 \\
OD-Ly-R & NOD-Ly-R & -1.27 & \textless 0.01 \\
OD-Ly-R & NOD-Ln-R & 1.58 & 0.05 \\
NOD-Ln-R & NOD-Ly-R & -3.75 & \textless 0.01 \\
NOD-Ly-R & MM-NOD-Ly-RGB & 1.38 & 0.02 \\
NOD-Ly-R & TN-NOD-Ly-RGB & -0.11 & 0.02 \\
MM-NOD-Ln-RGB & MM-NOD-Ly-RGB & -2.47 & \textless 0.01 \\
MM-NOD-Ly-RGB & MM-OD-Ly-RGB & 0.22 & 0.47 \\
MM-NOD-Ly-RGB & TN-NOD-Ly-RGB & -1.78 & \textless 0.01 \\
TN-NOD-Ln-RGB & TN-NOD-Ly-RGB & -0.48 & \textless 0.01 \\
TN-NOD-Ly-RGB & TN-OD-Ly-RGB & 0.01 & 0.14 \\
\end{tabular}
\end{ruledtabular}
\end{table}

The model selection based on the uncertainty of the calibration using the plane-based method was confirmed as a convenient approach to compare film dosimetry models, in the sense that lower uncertainties are correlated with higher expected $\gamma_{<1}$ values in the gamma analysis. Both model selection and expected gamma approaches produced the same conclusions. An additional validation was conducted by testing the statistical significance of the gamma analysis results. 

Models were compared using paired differences of $\gamma_{<1}$(3\% 3mm) values. The sample was composed of the 42 benchmark cases. Since the probability density functions of $\gamma_{<1}$ differences were not normal in general, the Wilcoxon signed-rank test was used. Thus, the null hypothesis was that the median difference between paired observations was zero. $P$-values of less than 0.05 were considered as statistically significant, {\it i.e.}, indicative that one of the models provided higher $\gamma_{<1}$(3\% 3mm) results. There are 990 possible pairwise comparisons with 44 models, for clarity only the 13 comparisons considered most relevant were studied. They are shown in Table~\ref{tab:inference}. It was found that applying lateral corrections significantly improved the gamma analysis results both for single-channel (R) and triple-channel models. Using NOD as film-scanner response improved the results, significantly for single-channel (R) but not significantly for triple-channel models. The NOD-Ly-R model was found to be significantly better than the MM-NOD-Ly-RGB model, and the most accurate model according to the model selection approach ({\it i.e.}, the TN-NOD-Ly-RGB model) also produced significantly higher $\gamma_{<1}$(3\% 3mm) values than both the MM-NOD-Ly-RGB and the NOD-Ly-R models.

\section{Conclusions}

Under the scope of applicability defined by the limits under which the models were tested ({\it i.e.}, the films were irradiated with megavoltage radiotherapy beams, with doses in the dose range of 20-600 cGy, entire films were scanned, the functional form of the sensitometric curves was a polynomial and the calibration followed the plane-based method \cite{mendez:2013}), it was confirmed that using TN perturbations provided better results than using MM perturbation models. It was found that applying lateral corrections produced more accurate film doses. The same occurred if the information of the non-irradiated film was incorporated by  scanning prior to irradiation, however this improvement did not produce significantly higher $\gamma_{<1}$(3\% 3mm) values with triple-channel models. Scanning prior to irradiation and applying lateral corrections had a greater effect in terms of improving the accuracy of the results than the increase of the number of combined color channels, which, for some models, did not necessarily improve the results.

Among the models under comparison, the most accurate was found to be the dosimetry model comprising TN perturbation, with NOD as response, applying the lateral correction, and employing all three color channels ({\it i.e.}, the TN-NOD-Ly-RGB model).

\begin{acknowledgments}
The author would like to thank Domingo Granero, Primo\v{z} Peterlin and Facundo Ballester for their very helpful comments during the preparation of this paper. 

The author is co-founder of Radiochromic.com. 
\end{acknowledgments}

\providecommand{\noopsort}[1]{}\providecommand{\singleletter}[1]{#1}%

\end{document}